\documentstyle[aps,prl,epsf,floats]{revtex}
\begin{document}
\twocolumn[\hsize\textwidth\columnwidth\hsize\csname@twocolumnfalse%
\endcsname
\title{Impurity in a $d$-wave superconductor: Kondo effect and STM
spectra}
\author{Anatoli Polkovnikov, Subir Sachdev, and  Matthias Vojta}
\address{Department of Physics, Yale University\\
P.O. Box 208120, New Haven, CT 06520-8120, USA}
\date{July 26, 2000}
\maketitle
\begin{abstract}
We present a theory for recent STM studies of Zn impurities in the
superconductor ${\rm Bi}_2 {\rm Sr}_2 {\rm CaCu}_2 {\rm O}_{8+
\delta}$, using insights from NMR experiments which show that
there is a net $S=1/2$ moment on the Cu ions near the Zn. We
argue that the Kondo spin dynamics of this moment is the origin
of the low bias peak in the differential conductance, rather than
a resonance in a purely potential scattering model. The spatial
and energy dependence of the STM spectra of our model can also
fit the experiments.

\end{abstract}
\pacs{PACS numbers:}
]

Recent progress \cite{seamus,ali} in scanning tunneling microscopy
(STM) of surfaces of the high temperature superconductor ${\rm
Bi}_2 {\rm Sr}_2 {\rm CaCu}_2 {\rm O}_{8+ \delta}$ has given us a
high resolution probe of its electronic correlations. Especially
notable have been studies\cite{seamus} in the vicinity of Zn
impurities on the Cu sites, and the site-specific information on
the variation in the pairing correlations. The ${\rm Zn}^{++}$
ion has spin $S=0$, and so it appears natural to interpret
\cite{seamus,ali} the experiments using a theoretical
model\cite{bala,tsu} in which each Zn site acts a potential
scatterer of Cu $3d$ electrons forming a $d$-wave BCS state.
However, several key experimental features do not fit easily in
this simple picture. There is a large peak in the differential
conductance close to zero bias near the Zn site: it is tempting
to identify this peak with a quasi-bound state in the potential
scattering model\cite{bala}, but such a state appears at low
energies only for a range of very large potential values
depending upon microscopic details \cite{alan,martin,ph}.
Furthermore, the spatial dependence of the zero bias peak is
unexpected \cite{henri}: it is largest on the Zn site, and with a
local maximum on the second neighbor Cu sites (see Fig~3c in
Ref~\onlinecite{seamus}), and this differs from the general
expectations of the potential scattering model \cite{bala,tsu}.
Finally, the observed spatially integrated spectrum is asymmetric
between positive and negative bias \cite{seamus}, while the
potential model predicts approximate symmetry \cite{flatte}.

We shall address these questions here using a rather different
theoretical model which was originally motivated by other
experiments on Zn or Li impurities (the ${\rm Li}^{+}$ ion is also
$S=0$). A series of beautiful NMR experiments
\cite{fink,alloul,bobroff,orsay1,orsay2,julien} have clearly shown
that each impurity, despite having no on-site spin, induces a
local, unpaired $S=1/2$ moment on the Cu ions in its vicinity at
intermediate energy scales. In the underdoped regime, this can be
understood\cite{fink,ssnr} by the confining property of the host
antiferromagnet, in which the impurity is a localized ``holon''
which binds the moment of a $S=1/2$ ``spinon''. At larger doping,
a related picture can be developed by analogy with the
theory\cite{hflm} for moment formation in the disordered metallic
state of Si:P---small variations in the potential combine with
strong local correlations to induce very localized spin
excitations. Other theoretical perspectives\cite{nl,fulde} on
local moment formation in the cuprates have also been given.
Further evidence for $S=1/2$ local moments near Zn sites appears
from neutron scattering experiments\cite{keimer}: it has been
argued\cite{science} that these are required to explain the
strong effects of a small concentration of Zn impurities on a
``resonance peak'' in the spin dynamic structure factor.

This paper will study a $S=1/2$ local moment near the Zn/Li site
coupled by exchange interactions to the fermionic $S=1/2$
excitations of a $d$-wave superconductor. We shall describe the
effects of the Kondo screening of the local moment, by the
fermionic excitations, on the STM spectra: we shall show that our
model leads naturally to a peak in the differential
conductance at a low bias of order the Kondo temperature, with a
spatial distribution which can fit the experiments. Our spatially
integrated spectrum also has a bias asymmetry which increases
with decreasing doping.

We begin by describing our model Hamiltonian, $H=H_{\rm
BCS}+H_{\rm imp}$, for a single Zn impurity \cite{science}. The first term
describes the host superconductor, which we model by a simple BCS
Hamiltonian
\begin{displaymath}
H_{\rm BCS} = \sum_{\bf k} \Psi^{\dagger}_{\bf k} \left[
(\varepsilon_{\bf k} - \mu) \tau^z + \Delta_{\bf k} \tau^x \right]
\Psi_{\bf k}.
\end{displaymath}
Here $\Psi_{\bf k} = (c_{{\bf k}\uparrow}, c_{-{\bf k}\downarrow}^{\dagger})$ is
a
Nambu spinor at momentum ${\bf k}=(k_x, k_y)$ ($c_{{\bf k}\alpha}$ annihilates
an electron with spin $\alpha$ on a 3d orbital),
$\tau^{x,y,z}$ are Pauli matrices in particle-hole space, and $\mu$ is
a chemical potential. For the kinetic energy, $\varepsilon_{\bf k}$ we
have first ($t$), second ($t^{\prime}$), and third
($t^{\prime\prime})$ neighbor hopping,
while we assume a
$d$-wave form for the BCS pairing function $\Delta_{\bf k} =
(\Delta_0/2)
(\cos k_x - \cos k_y )$. The Zn impurity is at
${\bf r} = {\bf r}_0 $, and is
described by
\begin{displaymath}
H_{\rm imp} = \sum_{{\bf r} \in {\cal N}} K({\bf r})
\vec{S} \cdot c^{\dagger}_{\alpha} ({\bf r})
\frac{\vec{\sigma}_{\alpha\beta}}{2} c_{\beta} ({\bf r})
+ U  c^{\dagger}_{\alpha} ({\bf r}_0) c_{\alpha} ({\bf r}_0),
\end{displaymath}
where $\sqrt{N_s} c_{\alpha} ({\bf r}) = \sum_k c_{{\bf k} \alpha}
e^{i {\bf k} \cdot {\bf r}}$ ($N_s$ is the number of sites in the
lattice) annihilates an electron at site ${\bf r}$, $\vec{\sigma}$
are the Pauli matrices in spin space, and ${\cal N}$ is a set of
sites in the neighborhood of ${\bf r}_0$. The spin degree of
freedom induced by the impurity is
represented by the $S=1/2$ operator $\vec{S}$. The
on-site potential scattering of the Zn ion is represented by $U$:
previous analyses of STM spectra of Zn ions\cite{bala,tsu,martin}
included only this second term in $H_{\rm imp}$ and
omitted\cite{flatte} the degrees of freedom represented by
$\vec{S}$. We assume that effects due to spatial variations
in the
self-consistent pairing amplitude near the impurity, along with
those due to
Hartree renormalizations from the
Coulomb interactions, have been absorbed into the
effective parameters $K({\bf r})$ and $U$ (as in \cite{hflm}); therefore, as
the dopant charge carriers are moving in a background of ions that
are both doubly charged (${\rm Cu}^{++}$ and ${\rm Zn}^{++}$), we
expect $U$ to be negative and measuring mainly the shift
in the $d$ level energy on the Zn (along the same
lines, $U$ should be larger and positive for ${\rm Li}^{+}$
impurities).

An important ingredient in our computation, which influences the
spatial form of the STM spectrum, is the ${\bf r}$ dependence of
the spin-dependent interaction $K({\bf r})$. This is quite
difficult to determine from first principles, and we will instead
use the spatial information obtained by analyses of Knight shifts
in NMR experiments \cite{orsay2,julien}. The $S=1/2$ moment is
found to be concentrated on the 4 Cu nearest neighbors of the Zn
ion (see Fig~1g in \cite{julien} and Fig~19 in \cite{orsay2}), and
the moment on the Zn site itself, ${\bf r} = {\bf r}_0$, is
negligible. We expect that the dopant holes will have a strong
preference to reside on a Cu site with a moment (as the Cu spins
on the other sites have paired with each other and gained
exchange energy); this attraction is realized by $K({\bf r})$,
and we take $K({\bf r}) = K_1$ for the 4 sites ${\bf r} -{\bf
r}_0 = (\pm 1,0), (0,\pm 1)$, while $K({\bf r}_0) = 0$. In
principle, it is not difficult to also include the smaller
$K({\bf r})$ values at larger values of $|{\bf r}-{\bf r}_0|$,
but we will neglect them here for simplicity. We also note that
while the above form for $K({\bf r})$ is reasonable, there is no
justification for a corresponding form for the potential
scattering term, which is surely largest at ${\bf r}={\bf r}_0$,
as in $H_{\rm imp}$.

The Kondo effect in $H$ is related to that in a class of models
which have been much studied recently \cite{withoff,chen,ingersent}:
these models
have a single spin coupled to a fermionic bath whose local density
of states, $\rho_{\ell} (\epsilon)$, vanishes as a power law
near the Fermi energy,
$\rho_{\ell} (\epsilon) \sim |\epsilon|^r$,
where $\epsilon$ is measured from the Fermi level.
The usual Kondo effect in a Fermi liquid corresponds to
$r=0$, while the present $d$-wave superconductor has $r=1$.
The value of $r$ and the presence or absence of
particle-hole symmetry \cite{ingersent} play a key role in
determining the low energy physics, and a comprehensive phase
diagram has been presented by Ingersent and collaborators \cite{ingersent}.
For $r= 1$
and with perfect particle-hole symmetry (for $H$ this
corresponds to $t^{\prime}=0$, $\mu=0$, and $U=0$) there is in
fact no Kondo effect: the spin is free at low energies for all
values of the exchange, $K$. However, there is a quantum
phase transition at a finite magnitude of particle-hole symmetry
breaking and at a finite $K$, to a phase in which the spin is Kondo
screened below an energy scale $T_K$.
The universal critical theory for this quantum critical
point is not known and we shall not discuss it here.

Here, we shall describe the dynamics of $H$ using a ``large $N$''
approach \cite{nick,hewson,withoff}.
While this method has numerous artifacts near the
quantum-critical point just mentioned, and is not quantitatively accurate,
it does capture the qualitative physics
of the Kondo screened phase in an effective manner, and this is
our primary interest.
We represent the spin $\vec{S}$ by a fermion, $f_{\alpha}$, which obeys the
constraint $f^{\dagger}_{\alpha} f_{\alpha}=1$. We impose this by
a Lagrange multiplier $\lambda$, and decouple the exchange
interactions by complex Hubbard-Stratonovich fields $\varphi_{\bf r}$;
$\lambda$ and $\varphi_{\bf r}$ are approximated by their static, real
saddle point values. The physical quantities are expressed in
terms of the Green's function of the Nambu spinor
$F= (f_{\uparrow}, f_{\downarrow}^{\dagger})$, which in
Matsubara frequency, $\omega_n$, is ${\cal T}(\omega_n)
= \langle F(\omega_n) F^{\dagger} (\omega_n) \rangle$.
We have
\begin{displaymath}
{\cal T}^{-1} (\omega_n)
= -i \omega_n + \lambda \tau^z -
\sum_{{\bf r},{\bf r}^{\prime} \in {\cal N}} \varphi_{\bf r}
\varphi_{\bf r^{\prime}}\tau^z G({\bf r},{\bf r}^{\prime},\omega_n) \tau^z
\end{displaymath}
where $G$ is the $\Psi$ Green's function with potential scattering
alone:
\begin{eqnarray}
G({\bf r}&&,{\bf r}^{\prime},\omega_n) =
G^0 ({\bf r}-{\bf r}^{\prime},\omega_n)
- U G^0 ({\bf r}-{\bf r}_0,\omega_n) \nonumber \\
&&\times \tau^z [ 1 +
U G^0 ((0,0),\omega_n) \tau^z ]^{-1} G^0 ({\bf r}_0-{\bf
r}^{\prime},\omega_n);
\end{eqnarray}
$G^0$ is the Green's function of the host $N_s G^0 ({\bf r}, \omega_n) =
\sum_{\bf k} e^{i {\bf k} \cdot {\bf r}} [-i \omega_n
+ (\varepsilon_k -\mu) \tau^z + \Delta_k \tau^x ]^{-1}$.
The values of $\lambda$ and $\varphi_{\bf r}$ were obtained by
numerically solving the constraint equations
$T\sum_{\omega_n} \mbox{Tr} [ \tau^z {\cal T}(\omega_n) ] = 0$
and
$\varphi_{\bf r}=-(T K ({\bf r})/2)\sum_{\omega_n} \sum_{{\bf r}^{\prime} \in
{\cal N}}
\varphi_{{\bf r}^{\prime}} \mbox{Tr}
[ \tau^z {\cal T}(\omega_n) \tau^z G({\bf r}^{\prime},{\bf r},\omega_n)
]$ ($T$ is the temperature, and Boltzmann's constant $k_B=1$).
Finally, the tunneling density of states (DOS) was obtained as
$\mbox{Im Tr}(\widetilde{G}(1+\tau^z))/2$, where $\widetilde{G}$ is
the full $\Psi$ Green's function
\begin{eqnarray}
&& \widetilde{G}({\bf r},{\bf r}^{\prime},\omega_n)
= G({\bf r},{\bf r}^{\prime},\omega_n) \nonumber \\
&+& \sum_{{\bf s},{\bf s}^{\prime}
\in {\cal N}}
\varphi_{\bf s} \varphi_{{\bf s}^{\prime}}
G({\bf r},{\bf s},\omega_n) \tau^z {\cal T}(\omega_n) \tau^z
G({\bf s}^{\prime},{\bf r}^{\prime},\omega_n).
\label{dos}
\end{eqnarray}

We now describe our numerical results. For small $K_1$, the
constraint equations have only the solution $\varphi_{\bf r}=0$.
In the present large $N$ approach, $\vec{S}$ is
completely decoupled
from the fermions at such a saddle point. This solution
corresponds to the free spin phase, and the STM spectra of the
large $N$ theory are
identical to that of the purely potential scattering model.
To include the spin dynamics, we take
$K_1 > K_{1c}$ below so that $\varphi_{\bf r} \neq 0$;
we found that the lowest free energy saddle points had a $d$-wave
pattern with $\varphi_{\bf r} = +[-]\varphi$
for ${\bf r} - {\bf r}_0 = (\pm 1, 0) [(0,\pm 1)]$.
Provided particle-hole symmetry is absent, the onset of a
non-zero $\varphi$ corresponds to a phase transition to
the Kondo screened phase at $K_{1c}$ (for the particle-hole symmetric case,
there is no Kondo screening \cite{ingersent}, and
the large $N$ equations show that this is so, even for $K_1 > K_{1c}$).
The assumption of eventual low energy Kondo screening is also in accord
with indications in NMR \cite{bobroff} and thermodynamic
measurements \cite{sisson}.
The large $N$ value for $K_{1c}$ is believed to be an overestimate:
this are other shortcomings will be
addressed in a forthcoming work employing an alternative
``non-crossing approximation'' \cite{hewson}.

For $K_1 > K_{1c}$, the energy dependence of the tunneling
spectrum is dominated by the form of the scattering matrix ${\cal
T}(\omega)$; the imaginary part of ${\cal
T}(\omega)$ shows a pronounced maximum at an energy $\omega_0$
(which becomes a very sharp peak near the Kondo transition),
so the location of possible peaks in the local
DOS is given by $\pm\omega_0$. On the other hand, $|\omega_0|$ can be
also be identified with the Kondo temperature, $T_K$. To
see this, consider the local impurity susceptiblity, which is
given by $\chi_{\text loc} = -(T/4) \sum_{\omega_n} \mbox{Tr} [ {\cal
T}(\omega_n)^2 ]$. Assuming the simple pole structure ${\cal
T}(\omega_n) = (-i \omega_n + \omega_0 \tau^z)^{-1}$ it is evident that
$\chi_{\text loc}$ changes its character at a $T$ of order
$|\omega_0 | \sim T_K$.

\begin{figure}[t]
\epsfxsize=2.9in \centerline{\epsffile{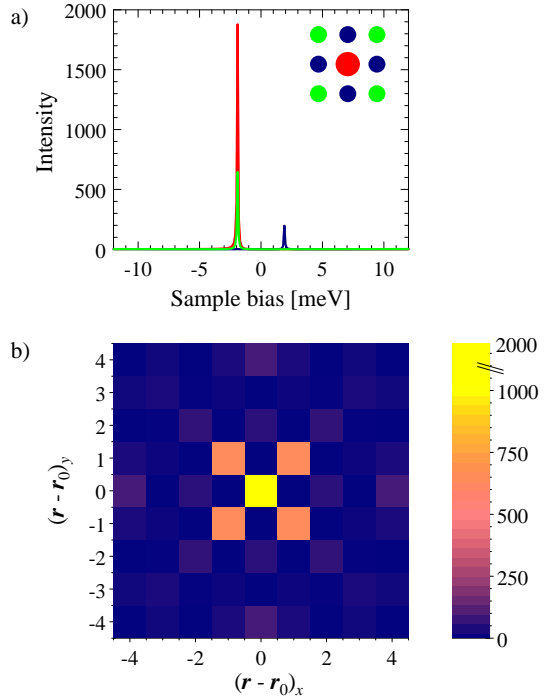}} \caption{
Calculated tunneling density of states for the Kondo impurity
model $H=H_{\rm BCS} + H_{\rm imp}$ at 14\% hole doping with a
realistic band structure ($t=0.1$ eV, $t^{\prime} = -t/4$,
$t^{\prime\prime} = t/8$), $\Delta_0 = 0.04$ eV, $K_1 = 0.21$ eV,
$\mu = -0.09$ eV and $U=-0.05$ eV. For these parameters $K_{1c} =
0.1$ eV. ({\em a}) Tunneling spectrum versus sample bias for the
impurity site (red) and the nearest (blue) and second (green)
neighbor sites. ({\em b}) Spatial dependence of the differential
conductance at the bias $\omega = -1.9$ meV ($=-\omega_0$). The
ratio of the spatial integrals at $\pm \omega$ is 1.5, and this
value increases with decreasing doping.} \label{fig1}
\end{figure}

\begin{figure}[t]
\epsfxsize=2.9in \centerline{\epsffile{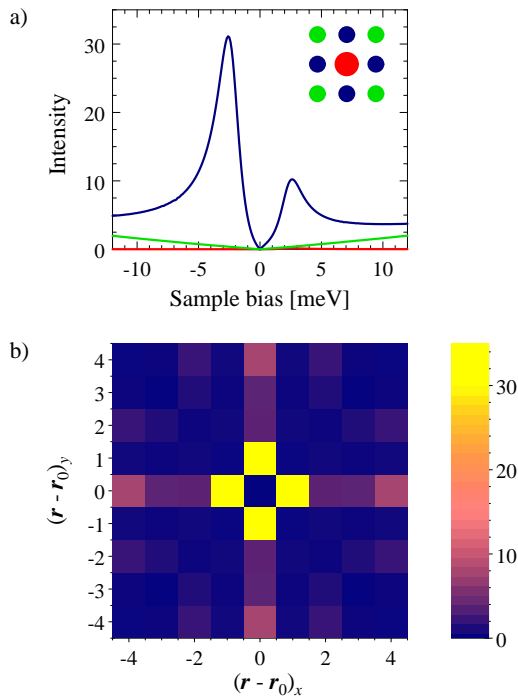}}
\caption{Parameters as in Fig~\ref{fig1}, but for a purely
potential scattering model ($H_{\rm imp}$ has $K_1 = 0$ and no
$\vec{S}$) with $U=-4 $ eV. The very large $|U|$ is necessary for
a low bias peak. The ratio of the spatial integrals at $\pm
\omega_0$ is now 1.05, and this value remains around unity with
decreasing doping.
}
\label{fig2}
\end{figure}

We show in Fig~\ref{fig1} the energy and spatial dependence of the
local DOS of $H$ for a typical set of parameters at 14\% hole
doping.
Most significant is the pronounced peak at ${\bf r}_0$ (on the Zn
site) at a bias, $\omega = -\omega_0$. This was a robust feature
of our results: the peak retained a small negative bias and width
for a wide range of $K_1
> K_{1c}$ and $U<0$. Away from the central site, the peaks were
much weaker, and varied slightly depending upon $U$ and doping:
for $-0.4$ eV $< U < 0$ at optimal doping (and more robustly at
lower doping) we obtained the alternating intensity pattern in
Fig~\ref{fig1}b \cite{stagger}. The spatial integral of the
spectrum at negative bias is larger than that at positive bias,
and the ratio is quoted in Fig~\ref{fig1}. For comparison we show
in Fig~\ref{fig2} the analogous results for a purely potential
scattering model \cite{bala} ($K_1 =0$ and $\vec{S}$ absent),
where we see dramatically different features. Now the largest
peak is on the first neighbors\cite{bala}; notice also the change
in the color scale in Fig~\ref{fig2}, and that this peak is not
as pronounced as the Zn site peak in Fig~\ref{fig1}. Further, we
had to choose a very large value of $|U|$ to make the peak sharp
and at a small negative bias. Both these features are sensitive
to variations in the values of the doping \cite{martin}, the band
structure ($t^{\prime}/t$, $t^{\prime\prime}/t$) and $U$: it is
not difficult to find broader peaks at higher energies, and to
switch the largest peak to positive bias.

We make a few remarks on the spatial dependencies in
Fig~\ref{fig1}. The dominant contribution at peak energy comes
from the second term on the right hand side of (\ref{dos}), and
both the normal ($\tau^z$) and anomalous ($\tau^x$) components of
$G$ are largest for sites on opposite sublattices. Using the
location of the four sites in ${\cal N}$ and the $d$-wave
structure of both the superconducting gap function $\Delta_{\bf
k}$ and the $\varphi_{\bf r}$ fields, we see that at ${\bf r}_0$
the normal (anomalous) Green's functions destructively
(constructively) interfere leading to a large peak at
$-\omega_0$. This interference is also responsible for the
asymmetry of the spatially integrated spectrum. (We note that for
an $s$-wave pattern in the $\varphi_{\bf r}$ fields the
interference mechanism is similar, with the difference that the
peak at ${\bf r}_0$ appears at $+\omega_0$.) An analogous
discussion\cite{bala} can be applied to Fig~\ref{fig2}, where
interference of scattering from different ${\bf r}$ is absent.
We also note that the coherence peaks, appearing in the DOS
at the gap energy, are much suppressed
at the impurity site and its neighbors, although we did {\em not}
include a spatial variation of the pairing amplitude.
Furthermore, upon moving to the metallic state by setting
$\Delta_0=0$, our approach is closely related to that of
\cite{freddy}, and then the large DOS around the Fermi level leads
to a Fano lineshape.

To summarize, this paper has introduced a new model ($H = H_{\rm
BCS} + H_{\rm imp}$) for STM spectra of Zn/Li impurities in the
high temperature superconductors. The model appears to be
required to explain NMR
\cite{fink,alloul,bobroff,orsay1,orsay2,julien} and neutron
scattering measurements \cite{keimer,science} on the same system.
It is therefore satisfying that our model also leads to a low
bias peak in the STM; the measured bias of this peak (1.5 meV)
suggests that the Kondo screening of the moment (as can be
measured directly in NMR) occurs at a temperature of order 15 K.

We conclude by noting related issues: ({\em i\/}) Our model
admits a simple generalization to Ni impurities. The Ni$^{++}$
ion has $S=1$, and so we add an additional $S=1$ degree of
freedom $\vec{S}_{\rm Ni}$. A simple choice for a Hamiltonian is
$H+ K^{\prime} \vec{S}_{\rm Ni}\cdot \vec{S}$. Depending upon the
values of $K^{\prime}$ and $K({\bf r})$, a rich variety of
behaviors appear possible. A likely possibility, suggested by the
NMR experiments \cite{orsay1}, is that $\vec{S}$ and
$\vec{S}_{\rm Ni}$ combine to form a $S=1/2$ moment; the Kondo
coupling of this effective moment is expected to be ferromagnetic
\cite{fulde}---so a $S=1/2$ moment remains unscreened. The STM
spectra will interpolate between the results described here, and
those in the presence of a static local magnetic field
\cite{nickel}, and this appears to be the case\cite{hudsonNi}.
({\em ii\/}) In a recent study \cite{science} of the broadening of
the collective spin resonance mode \cite{keimer} by Zn impurities,
the Kondo screening of $\vec{S}$ by the fermionic quasiparticles
was neglected. This is valid as long as the energy of the
resonance mode, $\Delta_{\rm res} \approx 40$ meV, is larger than
$T_K$; this condition is obeyed by the $T_K$ values discussed
above.

We thank H.~Alloul, W.~Atkinson, A.~Balatsky, S.~Davis,
G.~Khaliullin, A.~MacDonald, and S.~Pan for valuable discussions,
and US NSF Grant No DMR 96--23181 and the DFG (VO 794/1-1) for
support.


\begin{references}
\bibitem{ph} If the host band structure is particle-hole symmetric
(not the case at optimal doping), the potential
scattering ($U$ in $H_{\rm imp}$) has to be near
infinity
to obtain a low energy quasi-bound state, but in general, some finite
and non-universal value of $U$ is required.
\bibitem{stagger} We suspect that it is necessary to invoke the coupling to the
dynamic antiferromagnetic spin fluctuations \protect\cite{keimer,science}
to robustly explain the observed staggering \protect\cite{seamus}.
\end{references}
\end{document}